\documentclass[traditabstract]{aa} % for the abstract without structuration 
\usepackage{graphicx}
\usepackage{txfonts}
\begin{document}

\title{On the Raman O VI and related lines in classical novae}
\subtitle{{\it Letter to the Editor}}
\author{Steven N. Shore\inst{1,2}, Ivan De Gennaro Aquino\inst{1,3}, Simone Scaringi\inst{4,5}, Hans van Winckel\inst{4} }
\institute{
Dipartimento di Fisica "Enrico Fermi'', Universit\`a di Pisa; \email{shore@df.unipi.it}
\and 
INFN- Sezione Pisa, largo B. Pontecorvo 3, I-56127 Pisa, Italy
\and
Hamburger Sternwarte, Gojenbergsweg 112, D-21029 Hamburg, Germany
\and
Instituut voor Sterrenkunde, KU Leuven, Celestijnenlaan 200D, 3001 Leuven, Belgium
\and
 Max-Planck-Institut fuer extraterrestrische Physik, Giessenbachstra$\beta$e 1, 85741 Garching, Germany
 }

\date{Received ---; accepted ---}

\abstract{We critically examine the recent claimed detection of Raman scattered O VI at around 6830\AA\ in the iron curtain stage spectra of the classical CO nova V339 Del.  The observed line variations are compatible in profile and timing of emission line strength with an excited state transition of neutral carbon.  Line formation in classical nova ejecta is physically very different from that in symbiotic binaries, in which the O VI emission line is formed within the wind of the companion red giant at low differential velocity.  The ejecta velocity and density structure prevent the scattering from producing analogous features.  There might , however, be a broadband spectropolarimetric signature of the Raman process and also Rayleigh scattering at some stage in the expansion.  We show that the neutral carbon spectrum, hitherto under-exploited for novae,  is especially useful as a probe of the structure of the ejecta during the early, optically thick stages of the expansion.}

   \keywords{Stars, physical processes, novae }

  \titlerunning{Raman features in nova spectra} \authorrunning{S. N.
Shore et al.}
 \maketitle

\section{Introduction}%%%%%%%%%%%%%%%%%%%%%

In a recent paper, Skopal et al. (2014) have identified a line in the optical spectrum of V339 Del (Nova Del 2013) and other classical novae as the O VI Raman feature at around 6830\AA.  Raman scattering is an established radiative process in symbiotic stars.  It was first proposed by Nussbaumer et al. (1989) to explain the enigmatic 6825, 7083\AA\  emission line doublet. The success of the proposal was demonstrated by spectropolarimetric observations of these features, which is modulated with changing orbital phase (e.g. Schmid 1995, 2001).  The specific features are due to scattering of the O VI 1032, 1038\AA\ resonance doublet by Ly$\beta$.  In symbiotic systems, the O$^{+5}$ arises in a compact region around the white dwarf while the neutral hydrogen is much more extensively distributed in the wind of the companion red giant.  Besides O VI, other ions have been flagged as possibly producing Raman counterparts, especially the  H$\alpha$ wing and He II (e.g. Lee \& Lee 1997; Schmid 1995, 2001).  The lines are also variable on timescales of outbursts, correlated with the optical and UV variability of the white dwarf, and have a characteristic ratio that depends on the intrinsic variability of the O VI far ultraviolet doublet ratio.  Additional support for the mechanism comes from the observed conversion efficiencies that show a range of about 10 to 50 percent (see e.g. Schmid et al. 1999, Birrel et al. 2000).   Since in classical novae during the optically thick stage the column densities inferred for neutral hydrogen are as large as those derived for the red giant wind that forms the circum-dwarf environment, it would seem reasonable that such conditions could also produce the feature.  

This last consideration is behind the proposed identification of the Raman line in the  early spectra of V339 Del (Skopal et al. 2014).  The  optical maximum and immediately post-maximum stages of the nova outburst is due to recombination of the ejecta following the initial fireball (see Shore 2012).  Called the ``iron curtain'' (hereafter Fe curtain), the rapid increase in opacity due to the myriad accessible transitions of heavy neutral and singly ionized species produces a bolometric shift of the radiation from the central engine, the white dwarf, into the optical and near infrared, where the opacity is much lower.  If the white dwarf, at this early stage, has a surface temperature sufficiently high, the inner ejecta might be ionized to the O$^{+5}$ level.  The region will be, however, unobservable in either the X-ray or ultraviolet energy intervals because the column density of the neutral hydrogen is quite high, $> 10^{23}$ cm$^{-2}$ and the Fe curtain lines are very optically thick.  Regardless of the velocity or mass of the ejecta, this phase must occur, even if the ejecta are aspherical (e.g. Shore  2013).  Thus, the only way to probe the innermost portions of the ejecta is indirect.  In the Fe-group elements, transitions linked to the ground state that absorb between 4 and 10 eV serve as a sort of calorimeter, producing the observed permitted and forbidden lines and narrow absorption features during and after the recombination event whose velocities reflect the coupling of the individual transitions.  

For the Fe-group, this is extremely complicated.  There are so many individual transitions, densely spaced throughout the spectral interval, that identification of any one is a challenge.  For the resonance lines of the light elements, e.g., CNO, this is much easier.  If the white dwarf, which is undergoing CNO burning in the residual from the explosion, reaches an effective temperature $T_{eff} > 10^5$K, it can mimic the conditions in the symbiotics.  There is, however, an {\it essential} difference between these two extreme cases.  In symbiotic binaries, the red giant wind is slow and nearly comoving with the white dwarf.   As a probe of the winds, the closest analog to the usual suspects is the symbiotic-like recurrent novae. For V407 Cyg, high resolution observations during the first 100 days after the 2010 outburst did not show the Raman feature (e.g. Shore et al. 2011, 2012; Drozd et al. 2013).  It was detected in RS Oph from about Day 50 through 100 during the 2006 outburst by Iijima (2009) and also noted in the 1933 outburst by Joy \& Swings (1945) and in 1986 by Wallerstein \& Garvavich (1986) during the same interval.   In other systems, the limited spectroscopic coverage (e.g. V745 Sco 1989, 2014) prevents any comparison.  But the detection in RS Oph is quite similar to that of symbiotic binaries.  This has nothing to do with the ejecta.  It is purely from the illumination of the wind of the giant companion during the post-explosion supersoft phase when O VI is emitted around the relaxing white dwarf.  Nova ejecta are very different, as we now discuss.

\section{Why Raman features {\it cannot} form in nova ejecta}

The physical process for Raman scattering is absorption of into a virtual state in the Ly$\beta$ wing followed by the emission of the red line for each component of the doublet.  This leaves the hydrogen in the 2$s\ ^2$S state.   As explained for symbiotic binaries by Nussbaumer et al. (1989),  Schmid (1995, 2001), and Schmid et al. (1999), energy conservation requires that if $\nu_i$ is the incident line frequency (e.g. for each of the O VI components) and $\nu_{21}$ is the Ly$\alpha$ line frequency (the scattering ``target'' transition), then  $\nu_f=\nu_i-\nu_{21}$.  This also produces the polarized lines observed in the symbiotics at 6825\AA\ and 7083\AA.    The line width scales as $\Delta \nu_i/\nu_i = \Delta \nu_f/\nu_f$ so the scattered line should be broadened in wavelength by  $\Delta\lambda_f /\Delta \lambda_i = (\lambda_f/\lambda_i)^2 \approx 6.5$.  With this in mind, consider the case of nova ejecta.  The O$^{+5}$ region should be in the innermost parts of the ejecta where, if there is any FUV emission from the white dwarf it is completely absorbed, or from the hot atmosphere of the central star.  Neutral hydrogen is present over a much broader range of velocities but mainly in the outer parts of the ejecta where the recombination has already occurred after the end of the fireball stage.  Thus, the  Ly$\beta$ line is {\it redshifted} by about 1000 km s$^{-1}$ relative to the O VI doublet, or about +3.4\AA.  This should produce a {\it blueward} shift in wavelength of the Raman features of $\approx$-150\AA, to about 6670\AA .  Nussbaumer et al. (1989) list other resonance lines near Ly$\beta$ that might produce visible range Raman scattered features. This further complicates any identification in nova ejecta of any one such line.  For example,  O I 1028\AA\ produces a scattered line at about the same wavelength as the redshift effect described here for O VI.   A further complication is that the O VI are formed over a broad velocity range.  In later stages, spectra show that the UV resonance lines are also extremely broad (e.g. De Gennaro Aquino et al. 2014),  so the resulting line should be far broader than the usual observed width in the symbiotics.   We add that although the 7083\AA\ feature is not always visible in symbiotic binaries, it was definitely absent in V339 Del (see below for further discussion).

\begin{figure}[test]
  \begin{center}
  \includegraphics[width=9cm,angle=0]{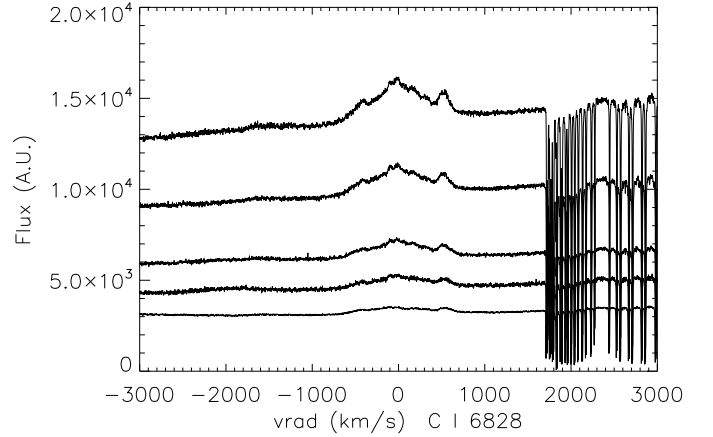}
  \end{center}
   \caption{Time sequence for C I 6828\AA.  Arbitrary scaling of the continuum.  From top to bottom: 2013 Aug. 25, Aug. 26, Aug. 30, Sep. 7, and Sep. 16.}
      \end{figure}

\begin{figure}[test]
  \begin{center}
  \includegraphics[width=9cm,angle=0]{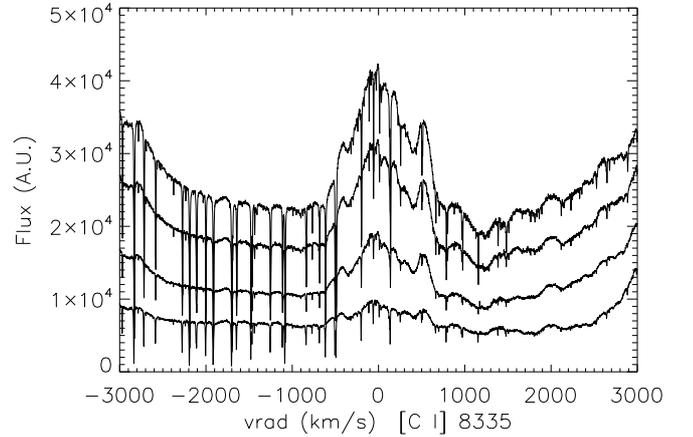}
  \end{center}
   \caption{Same sequence as Fig. 1, from 2013 Aug. 25 - Sep 7 for   C I 8335\AA\  (RMT 3F). }
      \end{figure}

\begin{figure}[test]
  \begin{center}
  \includegraphics[width=9cm,angle=0]{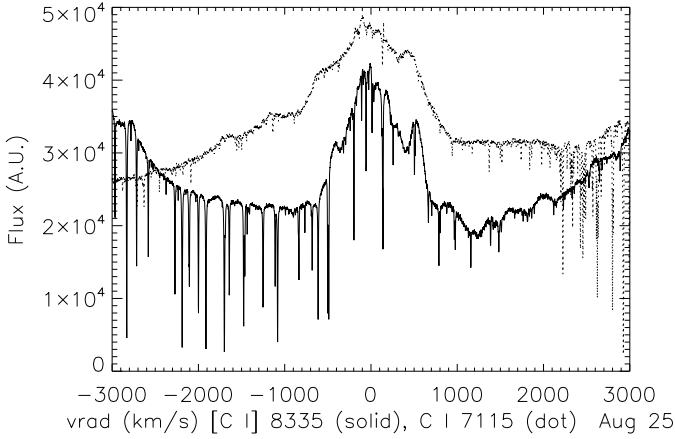}
  \end{center}
   \caption{Comparison between the  C I 8335\AA\ (solid) and C I 7115\AA\ (solid) line profiles on 2013 Aug. 25.  The flux units are arbitrary. }
      \end{figure}

\section{Alternate explanation: neutral carbon emission lines}

The emission line at 6828\AA\ is real.  If it is not a Raman feature, its persistence of the line during the optically thick stage and profile variations suggest an alternate interpretation.  

We observed V339 Del  with the {\it Hermes} spectrograph (Raskin et al. 2011), mounted on the 1.2 m Mercator Telescope at La Palma, Canary Islands, Spain. This highly efficient echelle spectrograph has a resolving power of R=86000 over the range 3800\AA\ to 9000\AA. The raw spectra were reduced with the instrument-specific pipeline.  The nova was observed on multiple occasions for over 3 months, beginning on 2013 Aug. 16.0UT  with regular daily monitoring during the first 14 days after its first detection, and every 2 to 3 days thereafter.  A time sequence of the 6828\AA\ feature, based on these data,  is shown in Fig. 1.  The identification appears to be C I 6828.12\AA\ (3p$^1$P$_1$-4d$^1$D$^o_2$), whose lower excitation is 68 856.33 cm$^{-1}$ (8.54 eV).  A significant clue is the line profile.  It is virtually the same as all other identifiable C I transitions in the 3800 - 9000\AA\ region of the V339 Del spectra  and also those of Fe II (which we take as a tracer low ionization species, see Fig. 4).  More important, the profile matches the  C I 8335\AA\ line quite closely (Fig. 2), if the C I identification is correct.  A comparison between the permitted C I 7115\AA\ and  C I 8335\AA\ profiles is shown in Fig. 3.   The [O I] 6300\AA\ line, which is present throughout the observational interval, is compared in Fig. 4 with the 6828\AA\ line.  As for  C I 8335\AA, the profiles closely match even in the fine structure on the emission.  The Fe II 6248\AA\ line, also shown in the figure, displays the same features.

\begin{figure}[test]
  \begin{center}
  \includegraphics[width=9cm,angle=0]{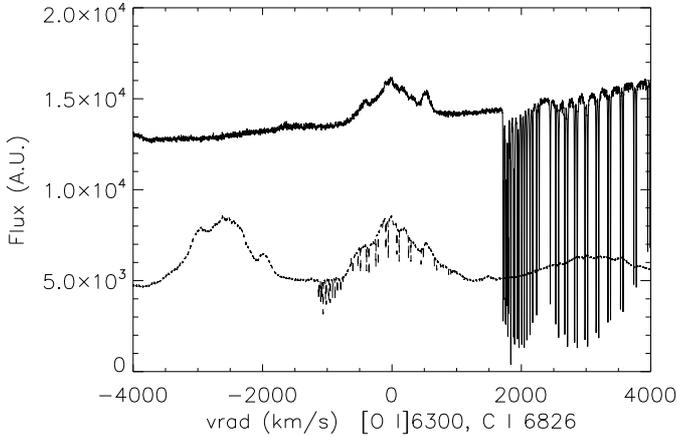}
  \end{center}
   \caption{Comparison between the [O I] 6300\AA\ (dotted) and C I 6828\AA\ (solid) line profiles on 2013 Aug. 20.  The strong blueward line next to [O I] is Fe II 6248\AA\ RMT 74}
      \end{figure}

The absorption feature, noted but unassigned by Skopal et al. (2014), matches the behavior of the other low ionization species at $v_{rad}\approx -600$ km s$^{-1}$ (e.g. Shore et al. 2013) and the same feature is seen on the  other C I transitions and the Fe II lines before about Aug. 20 (Fig. 5).  It also agrees with the low velocity absorption observed on Na I D1,D2 at the same epochs.  {\bf Furthermore, the Raman process has such a low cross section that an absorption is excluded}.   The absorption, which is then from an excited C I state, disappears and the line goes over into emission with the same structure as the other neutral lines.   In the spectra before Aug. 20, several excited state C I transitions showed a P Cyg absorption feature at around -600 km s$^{-1}$.  The lower state is populated by the resonance transitions around 1560\AA\ and the 1930\AA\ line during the optically thick stage, the upper states are populated by transitions at around 1193\AA.   The absorption feature, if due to the C I line, also matches the O I 7773\AA\ line and other excited state neutral species.  We note that Iijima (2012) identified a line in the recurrent nova CI Aql at 6830\AA\ as C I.\footnote{The 7773\AA\  emission line was also noted as a possible coronal species by Iijima (2009) in RS Oph but unidentified.  It could arise from the red giant wind as it recombines after shock breakout or by charge exchange, as in V407 Cyg (Shore et al. 2012).}   Other C I lines present included the composite features at 6014\AA\ and 7115\AA, 8335\AA, and 8727\AA.\footnote{ As noted by the referee, the 6828\AA\ and 8335\AA\ transitions are singlets while 7115\AA, being a triplet, should be broader.}

The feature on the blue wing of C I 7115\AA\ is likely C I 7087\AA\ blend, that could be mistaken for the Raman scattered O VI 1038\AA\ line.  For comparison, in Fig. 6 we show the variation of C II  4267\AA\ and note that C II 7235\AA\ also varies in the opposite sense to the 6828\AA\ line.

\begin{figure}[test]
  \begin{center}
  \includegraphics[width=9cm,angle=0]{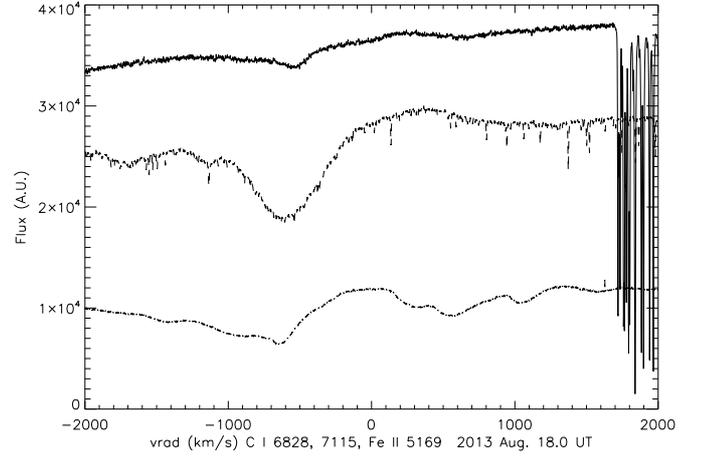}
  \end{center}
   \caption{Comparison between  C I 6828\AA\ (top, solid), C I 7115\AA\ (middle, dash), and Fe II 5169\AA\ during the absorption line stage of the 6828\AA\ profile.  The 5169\AA\ line is chosen to avoid possible blending with He I lines during the earliest stages of the expansion.}
         \end{figure}

\begin{figure}[test]
  \begin{center}
  \includegraphics[width=9cm,angle=0]{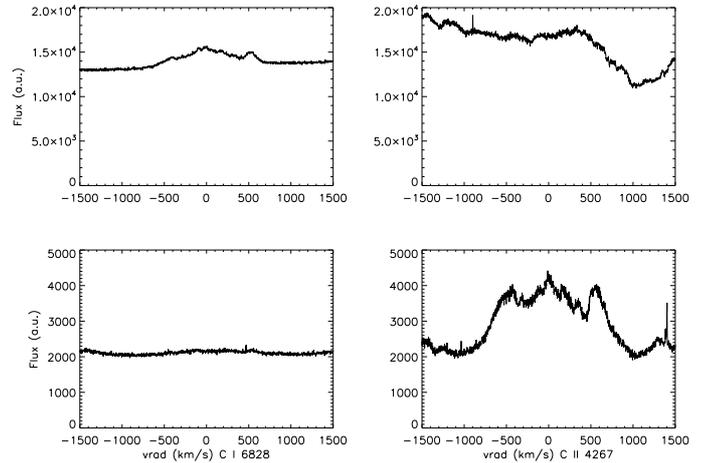}
  \end{center}
   \caption{Comparison between  C I 6828\AA\ (left column) and C II 4267\AA\ (right column) at two epochs.   Top: 2013 Aug. 24.9, bottom: 2013 Sep. 30.8.  Fluxes are in arbitrary units. }
      \end{figure}
 
To connect with the structure of the ejecta, we modeled the C I 7115\AA\  line with a bipolar structure for the ejecta (see  De Gennaro Aquino et al. (2014) and references therein).  The parameters for the model, shown in Fig. 7, are:  an outer angle $\theta_o$ = 20$^o$, inner angle $\theta_i$=90$^o$, and an inclination of about 55$^o$ to the line of sight. These are approximate, assuming an inverse cube (with radius) density law and a linear velocity law with v$_{max} \approx 2500$ km s$^{-1}$ and a radial thickness of about 0.7.  This implies that the C I is formed over a large part of the ejecta.  The constraints are weak from these profiles alone, acceptable inclinations ranged from 45$^o$ to 60$^o$ but spherical ejecta are ruled out.   This should emphasize that the ejecta are not unusual, similar results have already been found for many other classical and recurrent novae, and the appearance of the absorption on neutral carbon at the same time as the Na I and Fe II profiles argues for the onset (and later clearing) of the recombination wave in the ejecta (see e.g. Shore (2012) for a review).  The absence of {\it any} high ionization species, especially He II (which first appears only after Oct. 10) and any coronal lines (which do not appear before the end of 2013 Nov.) answers the conundrum posed by Skopal et al. (2014) about the peculiarity of O VI emission without accompanying high ionization species emission.  We also note that similar constraints are present in other high velocity gradient outflows.  A more complete analysis of the full multiwavelength spectral development is in preparation.\\

\begin{figure}[test]
  \begin{center}
  \includegraphics[width=9cm,angle=0]{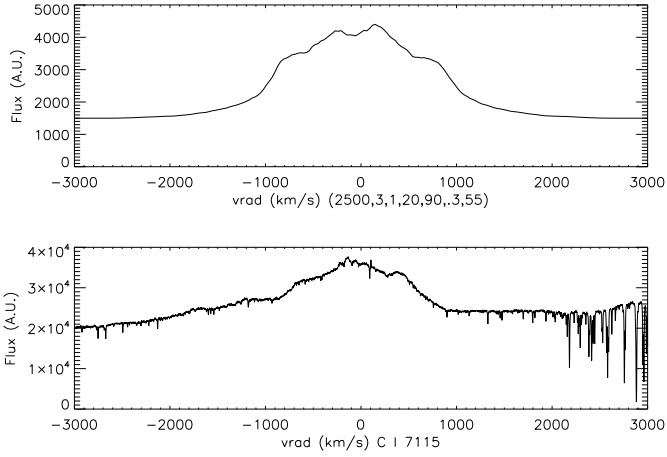}
  \end{center}
   \caption{Comparison between  C I 7115\AA\ (bottom) from Aug. 25 and a model profile.  See text for discussion.  The flux units are arbitrary. }
      \end{figure}

\acknowledgements Based on observations made with the Mercator Telescope, operated on the island of La Palma by the Flemish Community, at the Spanish Observatorio del Roque de los Muchachos of the Instituto de Astrof'sica de Canarias. Further based on observations obtained with the HERMES spectrograph, which is supported by the Fund for Scientific Research of Flanders (FWO), Belgium, the Research Council of K.U.Leuven, Belgium, the Fonds National de la Recherche Scientifique (F.R.S.-FNRS), Belgium, the Royal Observatory of Belgium, the Observatoire de Gen{\`e}ve, Switzerland and the Th\"uringer Landessternwarte Tautenburg, Germany.   SNS warmly thanks Elena Mason, Greg Schwarz, Sumner Starrfield, Francois Teyssier,  and Patrick Woudt for  invaluable discussions and exchanges.  SS acknowledges funding from the FWO Pegasus Marie Curie fellowship program. We  thank Alejandra Sans Fuentes, Roy Oestensen and Maria Suveges for carrying out some of the observations.  We thank the anonymous referee for very helpful suggestions.

\end{document}